\pdfoutput=1
\documentclass[conference]{IEEEtran}

\usepackage{graphicx}
\usepackage{booktabs}
\usepackage{amsmath}
\usepackage{amssymb}
\usepackage{multirow}
\usepackage{xurl}
\usepackage[table]{xcolor}
\usepackage{array}
\usepackage{cite}
\usepackage[hidelinks]{hyperref}

\title{MFSA: A Multi-Format Systolic Array with Native Micro Scaling Support}

\IEEEoverridecommandlockouts
\author{
\IEEEauthorblockN{Jiayi Wang\IEEEauthorrefmark{1}}
\IEEEauthorblockA{
\textit{University of Washington} \\
Seattle, WA, USA \\
jwang710@uw.edu
}
\and
\IEEEauthorblockN{Kearnan Bishop\IEEEauthorrefmark{1}}
\IEEEauthorblockA{
\textit{University of Washington} \\
Seattle, WA, USA \\
kearnan@uw.edu
}
\and
\IEEEauthorblockN{Ang Li}
\IEEEauthorblockA{
\textit{University of Washington} \\
Seattle, WA, USA \\
angliz@uw.edu
}
\thanks{\IEEEauthorrefmark{1}These authors contributed equally to this work.}
}

\begin{document}

\maketitle

\begin{abstract}
AI workloads increasingly employ low-precision formats such as
BF16 / FP16, FP8 variants, FP4, Micro Scaling (MX), and INT modes
across models, layers, and execution stages to balance accuracy,
memory footprint, and throughput, which in turn requires hardware to
support these formats in a flexible and efficient way. 
We present \textbf{MFSA}, a weight-stationary multi-format
systolic array that achieves area efficiency through two architectural
choices. At the PE level, MFSA uses dot-product-accumulator-based
processing elements extended from an existing open-source trans-precision FPU to support reduced-precision formats (FP32/FP16/BF16/CFP8/FP4/INT16/INT8/INT4) with a
shared datapath and proportional throughput scaling. At the array
level, MFSA employs \emph{Online Block-Scale Rebasing} (OBSR) for block-scaled formats, which folds each block scale into the running FP32 partial sum at block boundaries without adding a second cross-block accumulator, thereby reducing the additional overhead of block-scaling support. In
TSMC 28\,nm at 1\,GHz, matched-throughput BF16/FP8/FP4 cascaded
FMA-PE baselines are 1.5$\times$/2.5$\times$/4.5$\times$ larger than the
single-format DPA-PE, the gap growing with subword parallelism. A routed 16$\times$16 MFSA reaches a
peak 0.5/1/2/4\,TOPS of FP32/BF16/FP8/FP4, and
OBSR cuts the incremental MX overhead by 64.2\% in area and 67.5\% in
energy versus a naive MX implementation baseline, without hurting the numerical stability on real LLM traces. Adding support for all other formats based on a BF16+CFP8 systolic array costs 25.7\% additional area. OBSR also extends to non-power-of-two block scales: folding the fractional part of an E4M3 scale into the MFPE dot product supports NVFP4 alongside OCP MX for 11.8\% area and 7.3\% power over a CFP8+FP4 base, less than half the cost of a naive implementation.
\end{abstract}

\begin{IEEEkeywords}
Systolic array, multi-format arithmetic, fused dot product,
trans-precision, AI accelerator, deep learning hardware.
\end{IEEEkeywords}

\section{Introduction}\label{sec:intro}

\begin{figure}[t]
\centering
\includegraphics[width=0.9\columnwidth]{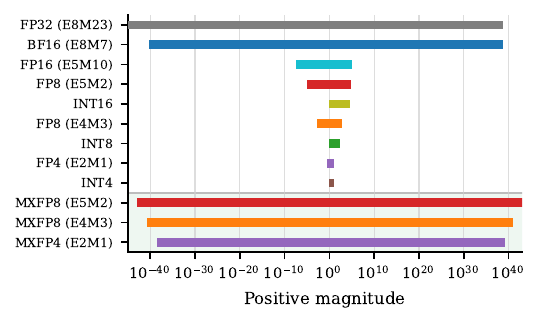}
\caption{Format encodings and positive magnitude ranges}
\label{fig:format_ranges}
\end{figure}

Modern AI workloads no longer operate at a single numeric precision.
While scientific machine learning models such as physics-informed
neural networks need high numerical precision, where reduced precision
degrades accuracy and stalls convergence~\cite{fp64pinn,sciml_mixed},
Large Language Models (LLMs) increasingly use lower-precision formats.
BF16 / FP16 remain common for training-state updates and optimizer
paths~\cite{bfloat16,kalamkar}, FP8 variants (E4M3/E5M2) target
high-throughput training and serving~\cite{fp8,h100}, FP4 is
emerging for aggressive inference~\cite{blackwell}, and Micro
Scaling (MX) formats pair low-precision elements with shared
block-scale exponents to recover dynamic range at low storage and
bandwidth cost~\cite{mx}. The preferred format can vary between
training and inference and even across layers within one model:
DeepSeek-V4, for example, mixes precisions even within attention and
Mixture-of-Experts (MoE) subsystems: KV entries keep BF16 on RoPE dimensions and FP8
elsewhere, the lightning indexer computes attention in FP4, and MoE
expert weights are quantized to FP4 and dequantized to FP8 for
computation~\cite{deepseekv4}.
Nemotron 3 Ultra's released checkpoint assigns precision per
operator~\cite{nemotron3ultra}: MoE routed experts in NVFP4, shared experts and Mamba~\cite{mamba} mixers in
FP8, attention and embeddings in BF16, with an FP8 KV
cache; Kimi K2 Thinking quantizes its experts
to INT4~\cite{kimik2}.
Lower precision reduces memory footprint and data-movement cost and
increases effective arithmetic throughput, but it also narrows dynamic
range and mantissa precision which potentially degrades performance, so format choice remains accuracy-sensitive~\cite{bfloat16,fp8}.
At the same time, accumulation needs to be at high precision to maintain numerical stability over the long reductions characteristic of AI matrix multiplication, so production systems retain FP32 accumulation for FP-input matmul~\cite{h100,blackwell,tpuv5p,trainium2,trainium3}.
Figure~\ref{fig:format_ranges} summarizes the representative
floating-point, integer, and MX modes targeted in this work and
highlights their markedly different magnitude ranges.

On the hardware side, commercial matrix engines have already shifted
toward this workload contract through explicit multi-format support,
even if their detailed processing element (PE) microarchitectures are not publicly described. Representative examples are summarized in
Table~\ref{tab:industry} using current public vendor documentation.
NVIDIA tensor cores~\cite{h100,blackwell} span BF16 / FP16 /
FP8 / FP32-class / INT modes, while Google TPU7x~\cite{tpu7x} and
AWS Trainium3~\cite{trainium3} expose matrix-engine throughput
across multiple low-precision modes together with explicit FP32-class
accumulation or output behavior. Together, these products define the
architectural target: a matrix-multiply engine that supports
multiple formats with roughly proportional throughput scaling.
However, the microarchitectural mechanisms needed to realize this
target efficiently, and the area/energy overheads of supporting
additional formats, remain under-characterized in the public
literature.

\begin{table}[t]
\centering
\caption{Representative commercial matrix engines: supported input formats and peak throughput
normalized to each product's BF16 mode ($1\times$).}
\label{tab:industry}
\scriptsize
\setlength{\tabcolsep}{2.2pt}
\renewcommand{\arraystretch}{1.0}
\begin{tabular}{@{}l c c c c c@{}}
\toprule
Format & H100 TC~\cite{h100} & B200 TC~\cite{blackwell} & TPU7x~\cite{tpu7x} & Trn3 (NCv4)~\cite{trainium3} & Dojo D1~\cite{dojo} \\
\midrule
FP32  & $0.5\times$ & $0.5\times$ & N/A       & $0.27\times$ & $0.06\times$ \\
BF16  & $1\times$ & $1\times$ & $1\times$ & $1\times$ & $1\times$ \\
FP16  & $1\times$ & $1\times$ & N/A       & $1\times$ & N/A \\
FP8   & $2\times$ & $2\times$ & $2\times$ & N/A       & $1\times$ \\
FP4   & N/A       & $4\times$ & N/A       & N/A       & N/A \\
MXFP8 & N/A       & $2\times$ & N/A       & $4\times$ & N/A \\
MXFP4 & N/A       & $4\times$ & N/A       & $4\times$ & N/A \\
INT8  & $2\times$ & $2\times$ & N/A       & N/A       & N/A \\
Accum.& FP32      & FP32      & FP32      & FP32/BF16 & FP32 \\
\bottomrule
\end{tabular}

\vspace{2pt}
\begin{minipage}{\columnwidth}
\scriptsize
TC = NVIDIA Tensor Core; TPU = Google Tensor Processing Unit; NCv4 =
4th-gen AWS NeuronCore (AWS values are the Tensor Engine inside
NeuronCore-v4). FP32 row = native FP32 matrix compute where available,
else FP32-class (e.g., TF32 on NVIDIA). Tesla Dojo uses \emph{configurable}
CFP8/CFP16; its CFP8 (runtime-selectable E4M3/E5M2) runs at the BF16
matrix rate, and FP32 is heavily deprioritized.
\end{minipage}
\end{table}

This work focuses on the microarchitectural design of such an
engine and on quantifying the overheads of supporting additional
formats. We use weight-stationary systolic matrix-multiply engines, as
exemplified by TPUs~\cite{tpuv1,tpuv4i,tpuv4}, because
they combine high data reuse with regular local connections and scale
naturally to large arrays. A simple way to support multiple formats is
to replicate a separate datapath for each format, but doing so
inflates area, routing, and control overhead while often leaving
hardware underutilized because workloads rarely exercise all formats
simultaneously. Achieving proportional throughput scaling on a shared
datapath therefore requires careful microarchitectural design. Beyond
supporting formats, MX formats introduce an
additional array-level overhead. In Open Compute Project (OCP) MXFP$\{8,6,4\}$~\cite{ocpmx}, each
32-element block carries a shared E8M0 scale, so a straightforward
systolic array whose PE spans more than one MX block needs to
cascade block-scaled partial sums and add a second FP32 state and
adder per block in each column. This work therefore targets an efficient and
flexible design that reduces these overheads and characterizes the
incremental cost of supporting additional formats.

To summarize, this work makes the following contributions:

\begin{itemize}
\item We propose the Multi-Format Systolic Array (MFSA), a
weight-stationary systolic array built around dot-product-accumulation (DPA)-based multi-format processing elements (MFPEs). The DPA unit is extended from the open-source TransDot~\cite{transdot} to
FP32/BF16/FP16/CFP8/FP4 and INT16/INT8/INT4 modes (CFP8 denotes
configurable FP8 with runtime-selectable E4M3/E5M2, as in Tesla
Dojo~\cite{dojo}). MFPE scales throughput through subword parallelism as
precision narrows, and preserves a wide FP32/INT32 accumulation path.

\item We employ \emph{Online Block-Scale Rebasing}
(OBSR) at block boundaries instead of adding a second cross-block FP32
accumulation path to support block-scaled formats. OBSR leverages the symmetry between the E8M0 MX scale and the FP32 exponent field, so cross-block rebasing reduces to small exponent-domain add/subtract operations. We further discuss methods to support the fractional NVFP4~\cite{nvfp4} (E4M3 block scale) and evaluate the area/power overhead.

\item We implement and evaluate MFSA in TSMC 28\,nm with a Cadence
Genus/Innovus/Joules flow. At the PE level, matched-throughput BF16/FP8/FP4 cascaded FMA-PE baselines are 1.5$\times$/2.5$\times$/4.5$\times$ larger than the single-format dot-product-accumulator PE, the gap growing with subword parallelism. Adding MX support through OBSR reduces the incremental MX area and energy overhead by 64.2\% and 67.5\% versus a
naive MX baseline while matching its numerics to within a few ULP on
real LLM traces. We further quantify the previously under-characterized
cost of incrementally supporting additional formats: a
full FP32/\allowbreak FP16/\allowbreak BF16/\allowbreak CFP8/\allowbreak FP4/\allowbreak INT16/\allowbreak INT8/\allowbreak INT4/\allowbreak MXFP8/\allowbreak MXFP4 array costs
only \textbf{25.7\%} more routed cell area than a BF16+CFP8 baseline array.
\end{itemize}

\section{MFSA Microarchitecture}\label{sec:pe}

\begin{figure*}[t]
\centering
\includegraphics[width=\linewidth]{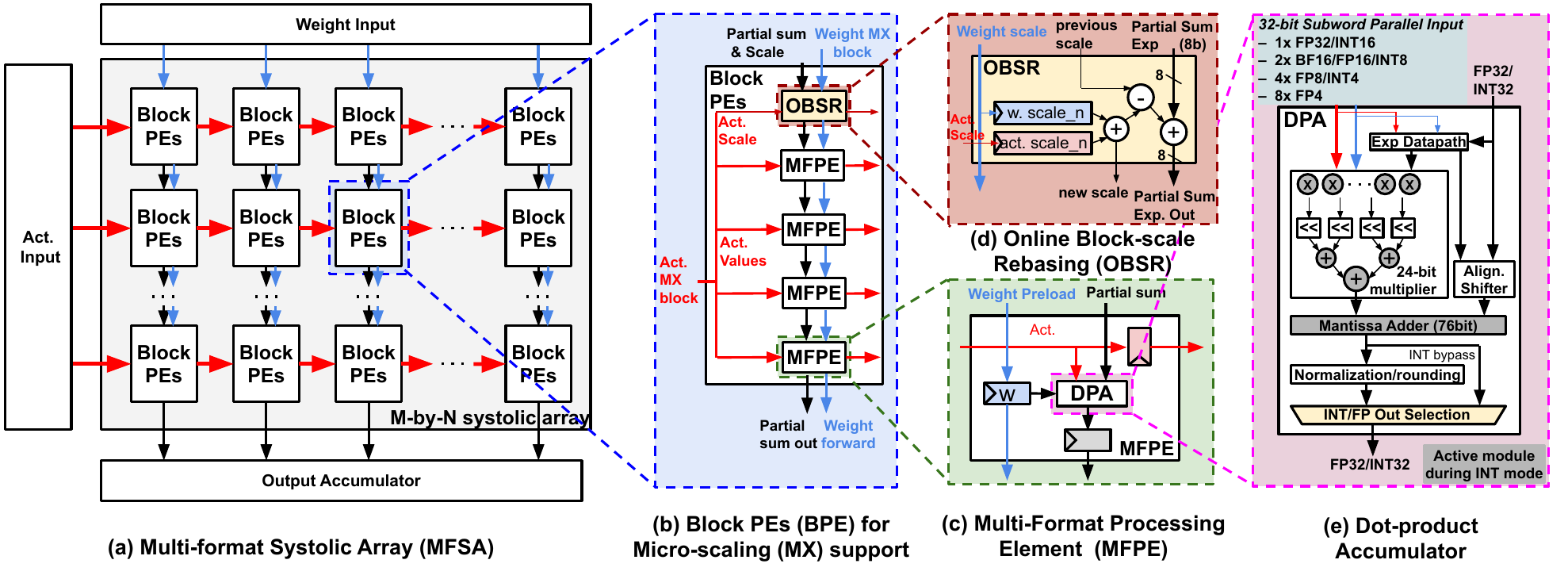}
\caption{MFSA hierarchy. (a) Top-level MFSA array. (b) For MX support,
four vertically adjacent MFPEs and one north-boundary OBSR unit form a
blocked PEs (BPE); MXFP4 uses one BPE per block, while MXFP8 spans
two stacked BPEs. (c) MFPE microarchitecture with DPA. (d) OBSR unit microarchitecture. (e) DPA inside each MFPE.}
\label{fig:pe}
\end{figure*}

\subsection{Top-Level Array Organization}

MFSA retains the classical weight-stationary systolic dataflow of
TPUs: an array of interconnected PEs that each perform the same
fused-multiply-accumulate (FMA) operation. Weights are preloaded and held locally inside each PE, activations stream west-to-east with the usual row
skew, and partial sums propagate north-to-south along a fixed FP32 (or
INT32) cascade.

Figure~\ref{fig:pe} details this organization. MFSA further groups four PEs
into a \emph{blocked PEs} (BPE) (Figure~\ref{fig:pe}(b)); each PE within a BPE is a
\emph{multi-format PE} (MFPE) (Figure~\ref{fig:pe}(c)). The arithmetic core of an MFPE is a
dot-product accumulator (DPA) (Figure~\ref{fig:pe}(e)) rather than a single FMA: it consumes subword-parallel reduced-precision operands and produces a
high-precision result, so throughput scales proportionally as precision
narrows. The array therefore preserves a mode-invariant partial-sum
path even as the input precision changes. Narrower formats travel as
naturally packed subwords on the same activation and weight links, so
MFSA needs no mode-specific inter-PE wiring or widened partial-sum
network for low-precision execution.

For MX modes, each BPE adds an Online Block Scale Rebasing (OBSR) unit (Figure~\ref{fig:pe}(d)) that performs cross-block rescaling before accumulating into the current block's results. 

A sideband format select signal configures the MFPE datapath per tile; because the north-south cascade stays FP32/INT32 in every mode, a format change never flushes the in-flight partial sums. Each column's FP32 output is then written to a result accumulator that combines partial sums across general matrix-multiply (GEMM) tiles.

The remainder of this section details each submodule.

\subsection{MFPE Datapath}

The targeted MFPE should support FP32\allowbreak /FP16\allowbreak /BF16\allowbreak /CFP8\allowbreak /FP4 inputs with FP32
accumulation and INT16/INT8/INT4 inputs with INT32 accumulation, while
delivering proportional throughput at each precision and high area
efficiency. Dot-product accumulation is a natural fit for this goal:
low-precision operands are packed into the original wide input datapath
and reduced by an internal dot product within each PE. This lets the
DPA share most of the wide FMA datapath across precision modes rather
than replicating a separate unit per format.

\paragraph{Floating-point datapath}
The open-source trans-precision dot-product unit of~\cite{transdot}
already supports FP32/FP16/FP8/FP4 with FP32 accumulation. Our DPA
extends it with the BF16, CFP8, and integer modes required by modern AI
models. The datapath comprises (1) an exponent-processing path; (2) a
mantissa-multiply path built around a reconfigurable $24{\times}24$
multi-mode multiplier that partitions into $2$, $4$, or $8$ narrower
products with their add trees; (3) an alignment shifter to align product mantissa and the mantissa of accumulated FP32 value; (4) a wide accumulator that sums the raw products and the mantissa of accumulated FP32 value; and (5) a normalization/rounding stage before the output.

The multi-mode multiplier produces $k$ products in carry-save form where $k$ is the number of terms in the dot-product operation, and the shared adder tree reduces them without intermediate rounding. The carry-save result is aligned against the FP32 accumulator's running exponent through a single shifter, and a 76-bit accumulator (24-b super-mantissa $+$ 2-b
guard/round/sticky $+$ FP32 dynamic-range alignment headroom) absorbs
the alignment with no precision loss within the cumulative shift range;
a single round-to-nearest-even (RNE) rounder produces the next FP32 partial sum. The MFPE therefore realizes one IEEE-754 normalization and RNE rounding on the summed $k$-product expression rather than $k$ separate normalization/roundings.

\paragraph{Integer extension}
To minimize the area cost of integer support, INT-$X$ reuses the
FP-$2X$ mantissa-multiplier slots: INT16 rides
the FP32 $24{\times}24$ multiplier ($k{=}1$); INT8 rides the
two-lane FP16-DP path ($k{=}2$); INT4 rides the four-lane FP8-DP
path ($k{=}4$, an exact fit since the E4M3 lane carries a 4-bit
significand: 3 stored mantissa bits plus the implicit bit). The multiplier
and accumulator are reused unchanged from the FP path; the only
INT-specific logic is integer sign-extension on the two's-complement
operands, signed/unsigned mode selection, and an output mux that
bypasses FP normalization and rounding.
Subword INT inherits the same 2$\times$-per-tier throughput cascade
as FP, at half the absolute rate of the same-width FP because INT-$X$
reuses FP-$2X$'s slots rather than FP-$X$'s. 

\subsection{Online Block-Scale Rebasing (OBSR) for MX}\label{ssec:obsr}
\paragraph{MX block arithmetic}
For OCP MX formats~\cite{ocpmx}, each operand vector is partitioned
into 32-element blocks, each carrying an 8-bit E8M0 shared exponent
$2^{s^{(b)}}$ on top of 32 per-element low-precision values $m^{(b,i)}$, each
itself an FP8/FP6/FP4 number with its own exponent and mantissa. The dot
product of two MX vectors $A$ and $B$ is
\begin{equation}
A\cdot B = \sum_{b} 2^{\,s_a^{(b)}+s_b^{(b)}}\sum_{i=0}^{31}
            m_a^{(b,i)}\, m_b^{(b,i)}.
\label{eq:mxdot}
\end{equation} 

The inner per-block product $\sum_i m_a^{(b,i)} m_b^{(b,i)}$ is already
computed by the DPA inside each MFPE; only the outer cross-block scale
combination remains, to be handled at the array level.

\paragraph{Naive systolic implementation baseline}
With each MFPE consuming $k$ elements/cycle (the dot-product width), an
MX block boundary falls every $\lceil 32/k\rceil$ vertically adjacent
MFPEs: the 4-MFPE BPE of Figure~\ref{fig:pe}(b) is the natural block
unit for MXFP4 ($k{=}8$), while MXFP8 ($k{=}4$) spans two stacked BPEs.
Because adjacent blocks carry different shared scales, their partial
sums cannot be added directly. A straightforward baseline implementation
therefore keeps \emph{two} FP32 values per column (an in-block
accumulator and a separately scaled cross-block running sum) plus a
dedicated FP32 cross-block adder at every block boundary. This second
FP32 accumulation path is the dominant overhead of MX support
(Section~\ref{ssec:eval_mxoverhead}).

\paragraph{OBSR: rebasing the running accumulator}
OBSR eliminates that second path by \emph{rebasing} the carried partial
sum: before a new block's products are accumulated, the inherited
accumulator is rescaled to the new block's combined scale, so the two
share a common scale and add directly on the existing FP32
cascade, with no second accumulator and no cross-block FP32 adder.

In general such a rescale is a divide and multiply by the scale ratio,
which is not free and can cost more than the second FP32 accumulation
path it replaces. OBSR is cheap precisely because MX block scales are E8M0,
i.e.\ \emph{pure powers of two}: rescaling by a power of two is an
integer add to the FP32 exponent field, leaving the mantissa untouched.

Concretely, OBSR maintains a \emph{single} running FP32 accumulator
$\tilde{S}$ and a signed 10-bit running combined-scale register
$s_{\text{cur}}$, together representing $V = 2^{s_{\text{cur}}}\tilde{S}$.
At each block boundary, the two incoming block scales first combine into
one power-of-two scale,
\begin{equation}
2^{s_{\text{new}}} = 2^{s_a^{(b{+}1)}}\cdot 2^{s_b^{(b{+}1)}}
\;\Longrightarrow\;
s_{\text{new}} = s_a^{(b{+}1)} + s_b^{(b{+}1)},
\label{eq:obsr_scale}
\end{equation}
the accumulator is rebased to this scale by perturbing only its
exponent field,
\begin{equation}
\tilde{S} \leftarrow \tilde{S}\cdot 2^{\,s_{\text{cur}}-s_{\text{new}}}
\;\equiv\;
e(\tilde{S}) \mathrel{+}= s_{\text{cur}}-s_{\text{new}},
\label{eq:obsr}
\end{equation}
where $e(\cdot)$ is the FP32 exponent field, and the running scale is
updated for the next block,
\begin{equation}
s_{\text{cur}} \leftarrow s_{\text{new}}.
\label{eq:obsr_scur}
\end{equation}
The boundary logic is thus integer-only: one add for the scale
combine (Equation~\eqref{eq:obsr_scale}), one subtract for
$s_{\text{cur}}-s_{\text{new}}$, and one add into the exponent
field (Equation~\eqref{eq:obsr}): \emph{two adds and one subtract}, mantissa
untouched. The next block's products then accumulate into the rebased
$\tilde{S}$ through the ordinary MFPE cascade.

\paragraph{Special values and MXFP6 recasting}
Each E8M0 scale is a power of two $2^{\,e-127}$ across the finite encodings
$\mathtt{0x00}$--$\mathtt{0xFE}$; $\mathtt{0xFF}$ is the reserved NaN code (no zero or infinity encoding), and a NaN scale propagates through the FP32 accumulator as a quiet NaN.
The rebase bypasses $\pm0$ and Inf/NaN, and clamps to the FP32 rails when the rebased exponent leaves range ($\pm\infty$ on overflow, flush-to-zero on underflow). Because OBSR holds $\tilde{S}=V\cdot2^{-s_{\text{cur}}}$, a block scale departing sharply from the accumulated history can clamp $\tilde{S}$ where a separate FP32 accumulator holding $V$ would not, though no such event occurs on real traces (Section~\ref{ssec:eval_obsr_robust}).
MXFP6 can be supported by on-the-fly  
recasting to MXFP8.

\begin{figure}[t]
\centering
\includegraphics[width=0.8\columnwidth]{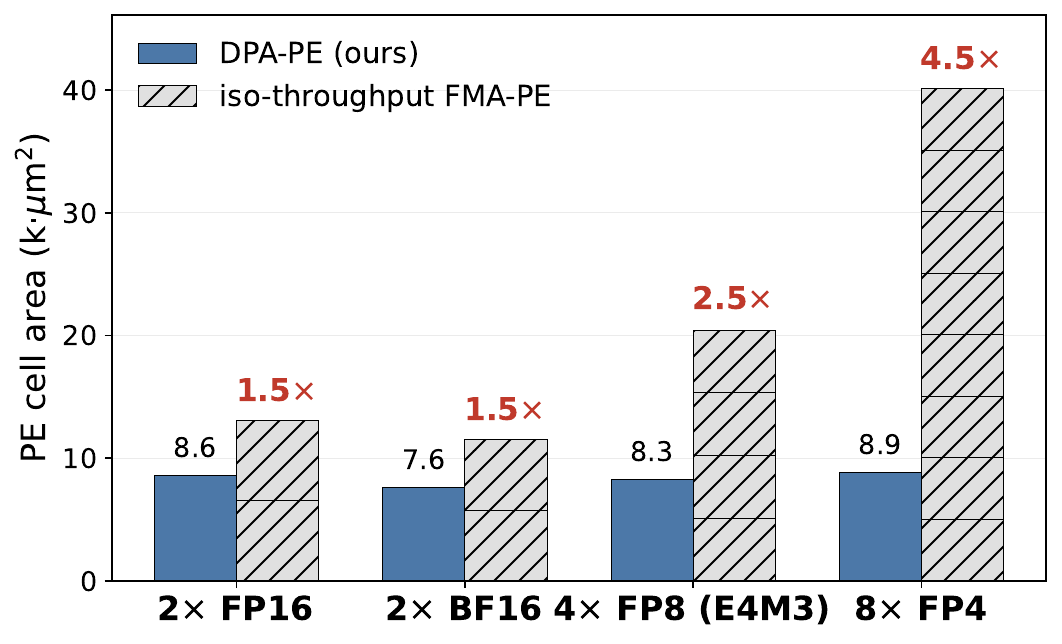}
\caption{Single-format DPA-PE vs.\ matched-throughput FMA-PE; the FMA
bar stacks $k$ cascaded single-format cells.}
\label{fig:dpa_fma}
\end{figure}

\section{Evaluation Methodology}\label{sec:method}
\subsection{Hardware implementation}
We implement MFSA in RTL and push it through synthesis with Cadence
Genus~\cite{genus} and the full place-and-route flow with Cadence
Innovus~\cite{innovus} in TSMC 28\,nm at the SS
$0.72$\,V/$125\,^\circ$C corner with a 1\,GHz target. Place-and-route uses
signoff RC extraction and automatic clock gating; the array routes at 59\% core utilization. Power is estimated with Cadence Joules~\cite{joules} driven by post-route parasitics and gate-level GEMM waveforms whose activations and weights are sampled from Qwen2.5-7B~\cite{qwen25} prefill traces (post-RMSNorm distributions rather than uniform random). We report peak steady-state power averaged over the activation-stream window,
excluding reset, weight-preload, and drain cycles.

\subsection{Baselines}
We compare MFSA against three baselines. (i) To isolate the PE-local
benefit of the DPA, we compare each single-format MFPE against a
\emph{matched-throughput} FMA PE built from $k$ cascaded single-stage
FMA units (e.g., a 2-term BF16 MFPE, latency 4, versus two cascaded
latency-2 BF16 FMA units, matching both throughput and latency).
(ii) To quantify OBSR, we compare a
$16{\times}16$ MFSA array with OBSR against the same array using a naive
MX baseline that maintains a second FP32 cascade word and a per-column
cross-block adder. (iii) To characterize the cost of incremental format
support, we generate MFSA RTL variants with different format masks,
each enabling a different format subset, and push each through
place-and-route for post-route area. 

\begin{figure}[t]
\centering
\includegraphics[width=\columnwidth]{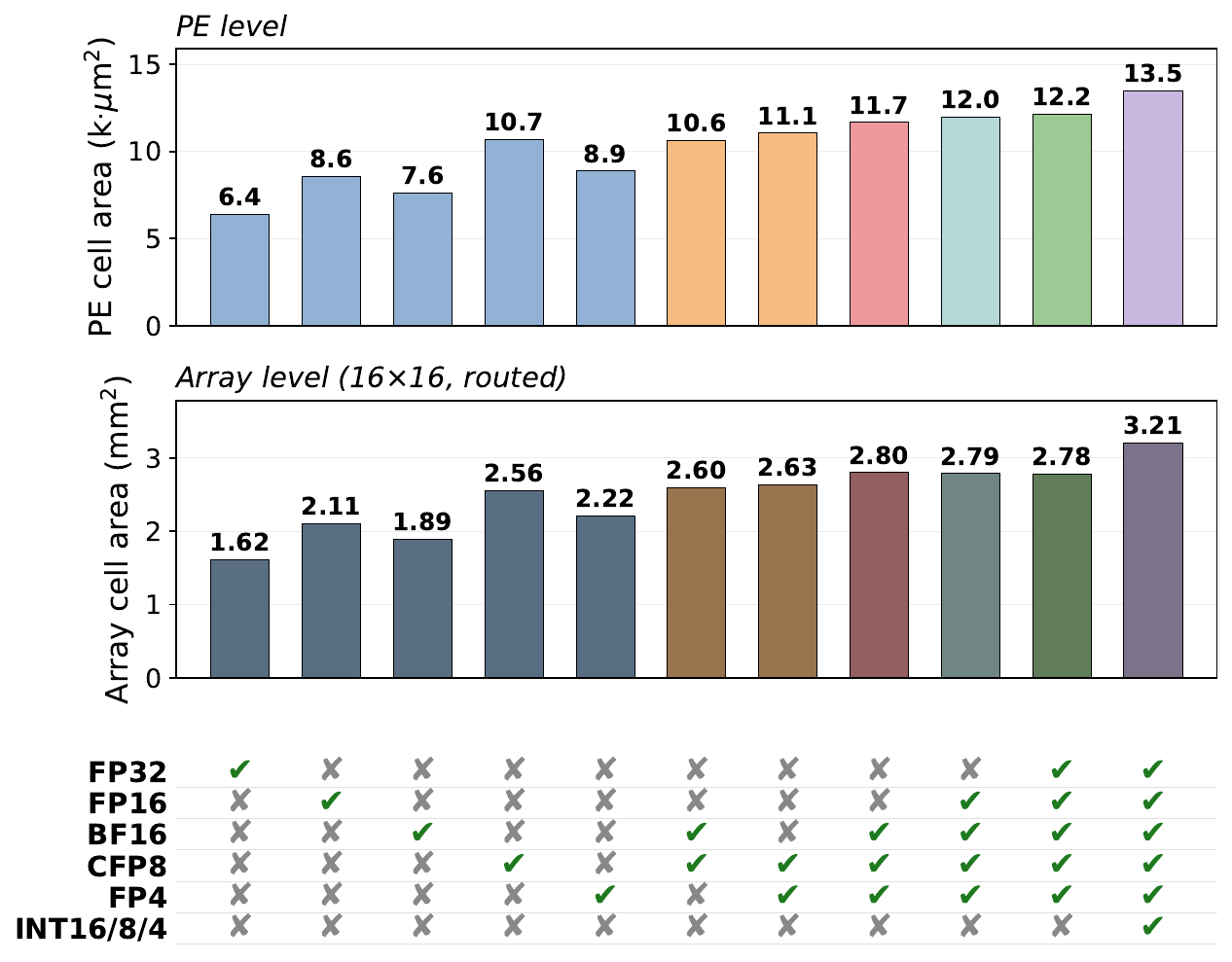}
\caption{PE and routed $16{\times}16$ array cell area across
enabled-format subsets.}
\label{fig:area_sweep}
\end{figure}

\begin{figure}[t]
\centering
\includegraphics[width=\columnwidth]{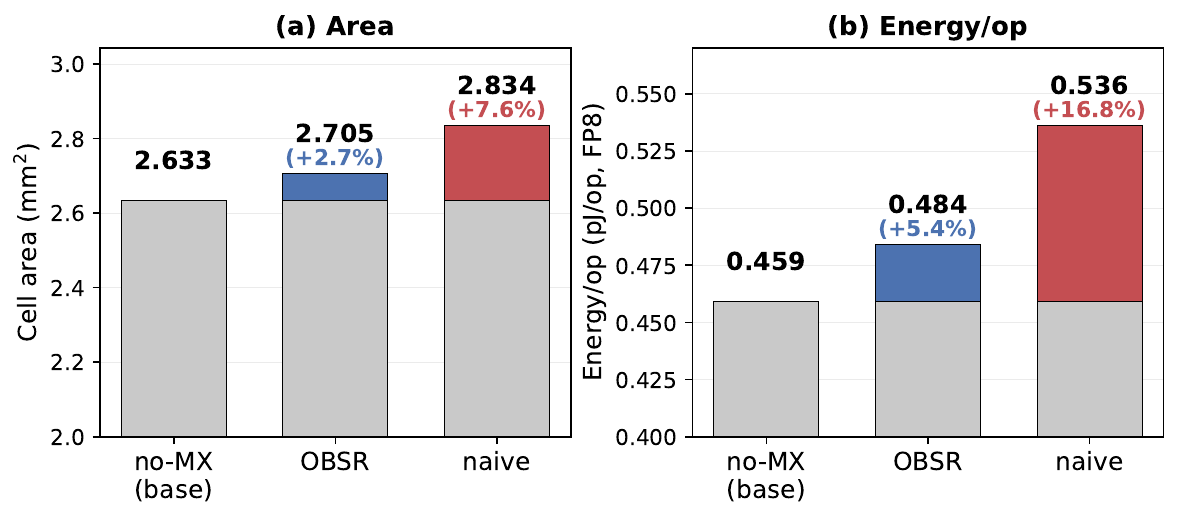}
\caption{Cost of MX support on the same 16$\times$16 CFP8/FP4
post-route array.}
\label{fig:mx_overhead}
\end{figure}

\begin{figure}[t]
\centering
\includegraphics[width=\columnwidth]{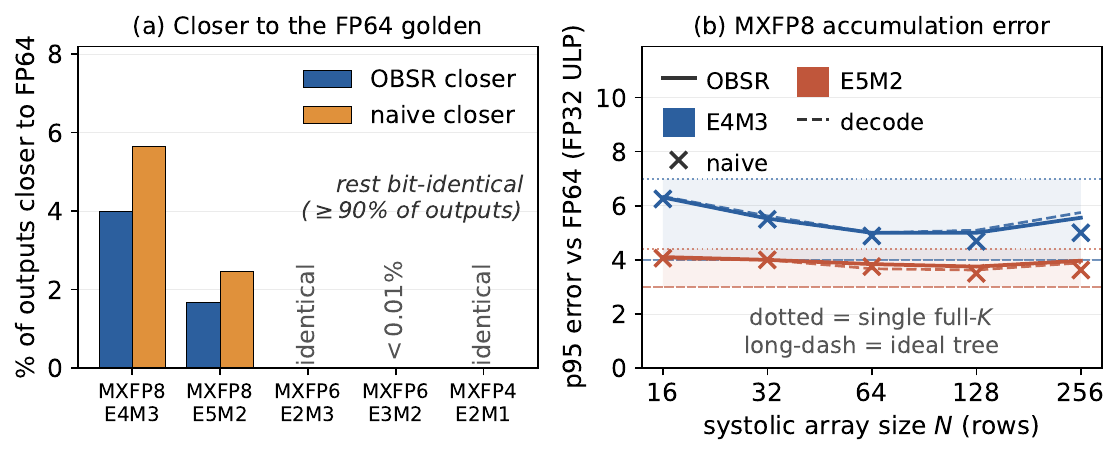}
\caption{OBSR robustness on real LLM activations (prefill and decode,
projection and attention GEMMs). (a) Share of outputs strictly closer to the
FP64 golden. (b) p95 error vs.\ array size for the two formats with non-zero
accumulation error; formats not shown are bit-identical.}
\label{fig:obsr_robust}
\end{figure}

\begin{figure}[t]
\centering
\begin{minipage}{0.46\columnwidth}\centering
\includegraphics[width=\linewidth]{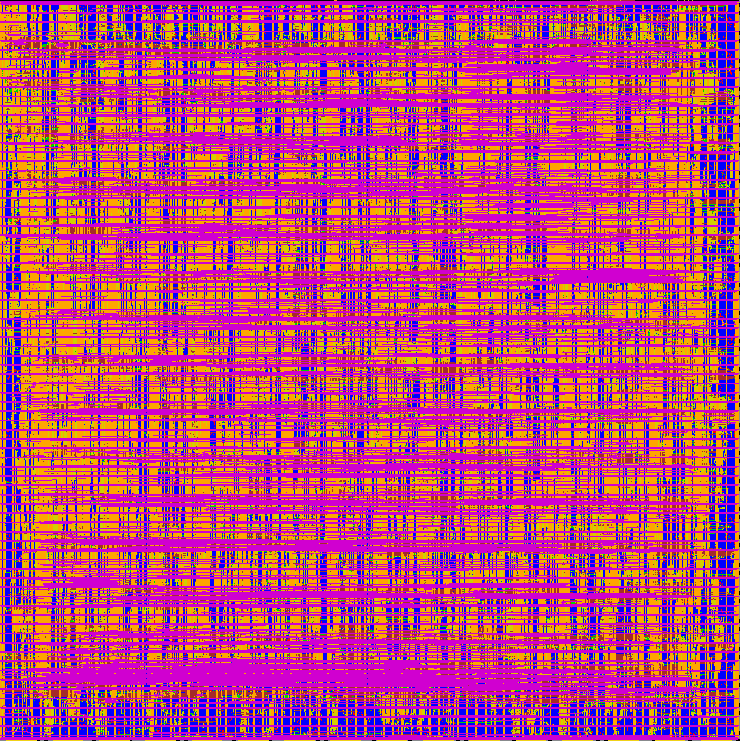}
{\footnotesize (a)}
\end{minipage}\hfill
\begin{minipage}{0.46\columnwidth}\centering
\includegraphics[width=\linewidth]{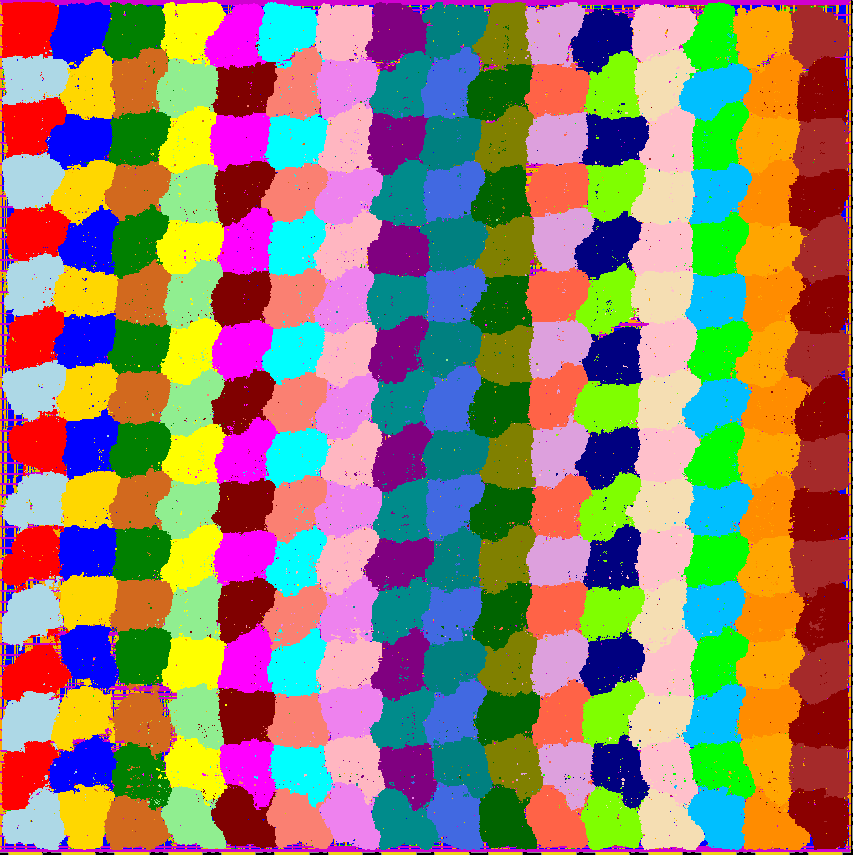}
{\footnotesize (b)}
\end{minipage}
\caption{16$\times$16 MFSA post-route layout in TSMC 28\,nm at 1\,GHz
(2.40$\times$2.40\,mm$^2$ die, 3.27\,mm$^2$ signoff cell area).
(a) Routed die. (b) The same die with each of the 256 MFPEs shaded.}
\label{fig:layout}
\end{figure}

\begin{table}[t]
\centering
\caption{Per-format throughput and energy of the full MFSA}
\label{tab:mxoverhead_full}
\footnotesize
\setlength{\tabcolsep}{6pt}
\begin{tabular}{@{}l l r r r@{}}
\toprule
Input Format & Acc. Format & Throughput & Power & Energy/op \\
       &    & (TOPS) & (W)   & (pJ/op)   \\
\midrule
FP32  &FP32  & 0.512 & 1.89 & 3.68 \\
FP16  &FP32  & 1.024 & 1.87 & 1.83 \\
BF16  &FP32  & 1.024 & 1.85 & 1.80 \\
CFP8  &FP32  & 2.048 & 1.85 & 0.90 \\
FP4   &FP32  & 4.096 & 1.85 & 0.45 \\
\midrule
INT16 &INT32  & 0.512 & 1.74 & 3.41 \\
INT8  &INT32  & 1.024 & 1.72 & 1.68 \\
INT4  &INT32  & 2.048 & 1.74 & 0.85 \\
\midrule
MXFP8 &FP32  & 2.048 & 1.85 & 0.90 \\
MXFP4 &FP32  & 4.096 & 1.85 & 0.45 \\
\bottomrule
\end{tabular}
\end{table}

\section{Evaluation Results}\label{sec:eval}

\subsection{MFPE area efficiency}\label{ssec:eval_pe_area}

\paragraph*{DPA-PE vs.\ FMA-PE}
Figure~\ref{fig:dpa_fma} compares each single-format DPA-PE against
matched-throughput, matched-latency cascaded FMA-PEs (Section~\ref{sec:method}):
both close 1\,GHz and produce one accumulated result per cycle, and the
FMA bar stacks its $k$ cascaded single-format cells. The DPA advantage
grows with subword parallelism:
at $k{=}2$ the cascaded FMA-PE is $1.5\times$ larger than the DPA-PE
(both FP16 and BF16), widening to $2.5\times$ at FP8 ($k{=}4$) and
$4.5\times$ at FP4 ($k{=}8$). This is the signature of deferred normalization and rounding:
amortizing a single IEEE-754 normalization and rounding over $k$ products saves more as $k$
grows, so the fused dot product structure pays off most in exactly the
low-precision regimes that dominate modern inference. 

\paragraph*{Cost of multi-format support}
Figure~\ref{fig:area_sweep} sweeps the enabled-format subsets in both PE-level and array-level; each
MFPE point is separately synthesized with the corresponding format
mask, and each array point is a full place-and-route. We report both PE
cell area and routed array area. On the routed array, broadening FP
support is cheap: relative to
a BF16+CFP8 array ($2.60$\,mm$^2$), enabling the full six-format FP set
(adding FP32, FP16, and FP4) costs only $+6.9\%$ ($2.78$\,mm$^2$), since
the shared trans-precision datapath is reused rather than replicated
per format. Adding INT16/8/4 on top costs a further $+15.4\%$
($3.21$\,mm$^2$) with the added integer sign-extension
and MUX logic. 

\subsection{OBSR vs. naive MX support baseline}\label{ssec:eval_mxoverhead}

Figure~\ref{fig:mx_overhead} reports the cost of \emph{provisioning} MX
support on the $16{\times}16$ CFP8+FP4 array, comparing OBSR against the
naive MX baseline (Section~\ref{ssec:obsr}). Over an MX-less version of
the same array, OBSR costs +2.7\% area and
+5.4\% energy/op, while the naive baseline costs roughly
2.5--3$\times$ as much (+7.6\% area,
+16.8\% energy/op). OBSR thus reduces the incremental MX cost by
64.2\% in area and 67.5\% in energy/op. On a full MFSA with all formats on, MX support adds only $+1.8\%$ more, giving the $3.27$\,mm$^2$ routed cell
area. Both overheads recur once per block of PE rows, so their ratio is set
per site rather than by array size; scaling to a $128{\times}128$ array
dilutes both slightly as fixed costs (clock tree, global signals) take a
larger share.

\subsection{OBSR numerical robustness}\label{ssec:eval_obsr_robust}
In a long floating-point reduction, numerical accuracy can depend on the order in which terms are accumulated. Each FP addition aligns operands according to their exponents; therefore, when a small term is added to a much larger running sum, low-order bits of the smaller operand may be rounded away. Although balanced pairwise summation often improves accuracy relative to purely sequential accumulation by reducing error growth, the most accurate order is data dependent, especially in the presence of cancellation or widely varying operand magnitudes.

This effect is relevant to our $N\times N$ array because the naive second-accumulator datapath and OBSR induce different FP32 addition orders. In the naive design, the 32 products within each inner MX block are first reduced to one FP32 partial sum. The resulting $N/32$ block sums are then accumulated sequentially by a second FP32 accumulator. In contrast, OBSR preserves the accumulation order of the original weight-stationary systolic array, which is equivalent to sequentially accumulating the $N$ products directly.

We quantify this effect on 168 GEMMs from prefill and decode
traces, covering both the projection layers (\texttt{down/up/o\_proj}, three
decoder depths) and the attention $QK^{\top}$ and $P{\cdot}V$ GEMMs, where
$P{\cdot}V$'s reduction grows with the context, at context lengths of 74 to
${\sim}430$ tokens, spanning four LLMs
(Qwen2.5-0.5B/1.5B/3B-Instruct and SmolLM2-1.7B-Instruct). FP32 outputs are
compared against an FP64 golden result accumulated in ascending-magnitude
order. OBSR and the naive baseline are bit-identical for MXFP4 and MXFP6
E2M3, whose block sums are exactly FP32-representable, and agree within a few
ULP for the rest (p95 within 6 ULP for MXFP8 on the projection GEMMs
and within 0.62 ULP on the attention GEMMs). Neither datapath is
consistently closer to the golden result (Fig.~\ref{fig:obsr_robust}(a)); the
naive method is closer slightly more often, as its block-local reduction lies
nearer to pairwise summation. Across array sizes from 16 to 256 rows
and across prefill and decode, the p95 curves nearly coincide
(Fig.~\ref{fig:obsr_robust}(b)), and large-ULP outliers arise only when
cancellation makes the FP64 reference near zero: the worst absolute error is
below $3\times10^{-5}$, at least $395\times$ smaller than the MX quantization
floor.

No clamp event occurs in the real-GEMM traces. Directed adversarial blocks (wide cross-block dynamic range,
alternating-sign cancellation, outliers, near-denormal scales) trigger 2242
clamp events across the four directed GEMMs and all array sizes. When the
clamp fires, the divergence is bounded by ${\sim}10^{-2}$ for MXFP4/MXFP6 but
reaches $8.19{\times}10^{3}$ for MXFP8 E4M3; OBSR thus differs from the
baseline only when a block scale departs far enough from the accumulated
history to clamp $\tilde{S}$, which never occurs in the measured traces.

A sweep over synthetic block-scale distributions (Gaussian, random-walk,
and alternating exponent patterns) shows 
OBSR tracks the naive baseline with zero clamp events until the block-scale
exponent spread reaches $\sigma\!\approx\!32$, at which point products span
FP32's full exponent range, while the measured LLM traces sit at
$\sigma\!\approx\!0.7$ -- $1.4$.

In summary, under real LLM workloads OBSR matches the naive
second-accumulator to within a sub-quantization residual while enabling the
area and energy savings of datapath reuse; the two schemes diverge only under
synthetic block-scale swings far beyond measured statistics. However,
in large-tile commercial engines (e.g.\ $128{\times}128$), where the tile far
exceeds the block size, a dedicated block-reduction path can also be a
deliberate accuracy-optimizing choice; the naive baseline models that
organization.

\subsection{Final PPA of the full multi-format MFSA + OBSR}\label{ssec:eval_full_ppa}

Table~\ref{tab:mxoverhead_full} summarizes the per-format power, performance, and area (PPA) of the implemented 16$\times$16 MFSA array, which supports the full format set: FP32, FP16, BF16, CFP8, FP4, INT16/8/4, and MXFP8/MXFP4. Figure~\ref{fig:layout} shows the post-route layout, a 2.40$\times$2.40\,mm$^2$ die with 3.27\,mm$^2$ of signoff cell area at 59\% core utilization. Each MFPE has a 4-cycle latency, and the array closes timing at 1\,GHz post-route. Across formats, peak power ranges from 1.72 to 1.89\,W under real LLM input traces. Energy per operation scales with throughput, decreasing from 3.68\,pJ/op for FP32 to 0.45\,pJ/op for FP4.

\section{Related Work and Comparison}\label{sec:related}

\begin{table}[t]
\centering
\caption{MX-format PPA of the full MFSA and Vmxdotp}
\label{tab:related_mx}
\footnotesize
\setlength{\tabcolsep}{2.5pt}
\renewcommand{\arraystretch}{1.1}
\begin{tabular}{@{}l c c *{6}{c}@{}}
\toprule
Work & Area & Power & \multicolumn{6}{c}{Format} \\
\cmidrule(lr){4-9}
 & (mm$^2$) & (W)
 & \multicolumn{3}{c}{MXFP8} & \multicolumn{3}{c}{MXFP4} \\
\cmidrule(lr){4-6}\cmidrule(lr){7-9}
 & & & T/A & T/W & pJ & T/A & T/W & pJ \\
\midrule
\rowcolor{gray!12}
\textbf{MFSA (ours)}, 28\,nm & 3.27 & 1.85
 & 0.63 & 1.11 & 0.90 & 1.25 & 2.21 & 0.45 \\
Vmxdotp~\cite{vmxdotp} $^\ddagger$, 12\,nm & 0.18 & 0.17$^\S$
 & 0.69 & 0.75 & 1.33 & 1.37 & 1.57 & 0.64 \\
Vmxdotp~\cite{vmxdotp} $^\ddagger$, 28\,nm$^\dagger$ & 0.99 & 0.40
 & 0.13 & 0.32 & 3.10 & 0.25 & 0.67 & 1.49 \\
\bottomrule
\end{tabular}

\vspace{2pt}
{\footnotesize\raggedright
T/A $=$ TOPS/mm$^2$, T/W $=$ TOPS/W, pJ $=$ pJ/op; both designs run at
1\,GHz; MFSA row is the routed $16{\times}16$ array (MX-mode power).
$^\ddagger$\,FPU only;
$^\S$\,power derived from the reported efficiency.
$^\dagger$\,first-order node scaling.\par}
\end{table}

\subsection{Trans-precision floating-point units}
Unit-level trans-precision FPUs and fused dot-product units, such as
FPnew~\cite{fpnew}, trans-precision FP units~\cite{transprec1},
TransDot~\cite{transdot}, and ExSdotp~\cite{exsdotp,miniflotats},
established the shared-datapath template for multi-format arithmetic
with deferred rounding and FP32 accumulation. MFSA's MFPE extends this
line to a more complete format set, supporting BF16, CFP8, and
INT16/8/4 on one datapath with compile-time and runtime configuration.

\subsection{Block floating point and microscaling}
Sharing one exponent across a block of mantissas is long-standing, from
classical block floating point to Flexpoint~\cite{flexpoint} and Microsoft
Floating Point~\cite{msfp}; OCP MX~\cite{ocpmx} is a fine-grained instance,
pairing 32-element blocks of low-precision elements with an E8M0 shared scale.
Such formats keep a \emph{separate} block-shared exponent and align mantissas
to it before an order-independent reduction. Recent MX hardware follows this
pattern at unit or cluster scope: Vmxdotp~\cite{vmxdotp} and
MXDOTP~\cite{mxdotp} add per-block MX dot products to RISC-V cores, a
precision-scalable datapath targets edge robotics~\cite{mxrobotics}, and
VEGETA~\cite{vegetamx} multicasts block scales into a sparse/dense GEMM array,
each applying the scale at the multiplier inputs and replicating scale handling
across lanes. OBSR instead folds each block scale into the exponent of a live
FP32 running accumulator, in place and mid-reduction, so MFSA carries MX
natively through a large weight-stationary systolic array with no separate
block exponent, no per-PE scale logic, and no second cross-block accumulator.
Table~\ref{tab:related_mx} compares MFSA's MX modes against
Vmxdotp~\cite{vmxdotp}. 

\subsection{Systolic and precision-scalable arrays}
Systolic and precision-scalable arrays, including TPU/Trainium, Tesla
Dojo~\cite{dojo}, RaPiD~\cite{rapid}, Bit
Fusion~\cite{bitfusion}, and BitBlade~\cite{bitblade}, explore
the relevant spatial organizations. RaPiD uses separate datapaths for
FP and INT rather than sharing hardware; the commercial parts likely
combine formats, but their microarchitectures are unreleased and their multi-format overheads are uncharacterized. Moreover, their reported results include chip-level components such as SRAMs, control logic, networks-on-chip, and memory controllers, which fall outside the scope of this work and make direct comparison inappropriate. We therefore do not compare against them. MFSA keeps one arithmetic substrate across FP, INT, and MX modes and quantifies the incremental cost of each. The contribution is therefore architectural: starting from a prior
multi-format FDP PE, MFSA shows how to integrate that PE into a routed
weight-stationary systolic array with mode-invariant partial-sum
transport and low-overhead MX support. 

\section{Discussion}\label{sec:discussion}
\subsection{Fractional block scales (NVFP4)}
\label{ssec:eval_nvfp4}
OBSR exploits E8M0 block scales: a power-of-two scale folds into the
accumulator exponent as an exact integer add. NVFP4~\cite{nvfp4} instead
pairs 16-element blocks with fractional E4M3 scales to recover model accuracy
at 4-bit precision, and applying such a scale wholesale would require the FP
multiplier per boundary that OBSR cannot support directly.

A method to support the fractional scale with OBSR is to disaggregate the scale's exponent from its mantissa, $s = 2^{e}\,(1.m)$. The power-of-two half $2^{e_a+e_b}$ follows the OBSR path
unchanged. The fractional half $\mu = (1.m_a)(1.m_b)$, an 8-bit integer, is
folded into the FP4 dot product \emph{inside} the MFPE: the four lane
magnitudes are summed first, then scaled by $\mu$ with a single
$11{\times}8$ multiply before the FP32 addend join. By distributivity the
fold is exact with no rounding and no divider, and the scale cascade keeps its
E8M0 width. 

Post-route, supporting OCP MX and NVFP4 costs an additional $+11.8\%$ cell
area and $+7.3\%$ power with OBSR$+\mu$. The naive method, which scales back
with an additional multiplier and a cross-block FP32 adder, incurs $+25.1\%$
area and $+24.0\%$ power. We further tested a second alternative that scales
back the product in every PE, avoiding the cross-block adder and OBSR
entirely but adding a multiplier and a dynamic exponent add per PE: it
incurs $+15.6\%$ area and $+15.7\%$ power.

Among these methods, OBSR$+\mu$ remains the most efficient, cutting the
incremental overhead by $24\%$ in area and $53\%$ in power versus the
second-best method (per-PE scale-back), and by $53\%$/$70\%$ versus the
dual-accumulator.

\subsection{Integer-heavy edge inference}
The integer formats ride the FPU's multipliers and thus reach half the
rate of the same-width FP formats (INT8 tracks BF16/FP16 rather than CFP8).
A model dominated by integer operations can add a dedicated integer path to
double integer throughput. Adding an additional INT dot-product path to match FP's throughput incurs additional
$1.39$\,k\,$\mu$m$^2$ ($+10.3\%$) per PE in the same setup.

\section{Conclusion}\label{sec:conc}
MFSA integrates a shared datapath dot-product accumulator (DPA) as the
leaf cell of a routed weight-stationary systolic array, keeping a single
shared datapath across FP32, FP16, BF16, CFP8, FP4, INT16/8/4, and the
MX block-scaled modes. Two choices make this efficient: a shared datapath DPA-based MFPE and Online
Block-Scale Rebasing mechanism, which folds the MX block scale into the running
FP32 accumulator rather than a second cross-block accumulator.
In a routed 16$\times$16 array at 1\,GHz in TSMC 28\,nm, the
matched-throughput cascaded FMA-PEs are 1.5--4.5$\times$ larger than the
single-format DPA-PE from BF16 to FP4, and OBSR cuts the incremental MX overhead by 64.2\% in area and 67.5\% in
energy with numerics matching the naive
baseline to within a few ULP on real LLM GEMMs. Adding support for all other formats on a BF16+CFP8 systolic array costs 25.7\% additional area. OBSR is not confined to power-of-two scales: disaggregating an E4M3 scale into an exponent that follows the OBSR path and an 8-bit mantissa folded into the MFPE dot product supports NVFP4 alongside OCP MX for 11.8\% area and 7.3\% power, roughly half of a naive cross-block accumulator. Together these make MFSA a compact multi-format substrate.

\bibliographystyle{IEEEtran}
\bibliography{references}

\end{document}